\documentclass[12pt,preprint,a4paper]{aastex}
\usepackage{times}
\usepackage{graphics,epsf}
\usepackage{amsmath}                % American Mathematical Society package
\usepackage{amsfonts}               % American Mathematical Society fonts
\usepackage{amssymb}                % American Mathematical Society symbol
\usepackage{epsfig}                 % EPS figures
\usepackage{graphicx}
\usepackage{url}
\usepackage{tabularx}
\usepackage{booktabs}
\def \s{~\rm{s}}
\def \km{~\rm{km}}

\def \AU{~\rm{AU}}
\def \erg{~\rm{erg}}

\def \yr{~\rm{yr}}

\begin{document}

\title{Planetary nebula progenitors that swallow binary systems}

\author{ Noam Soker\altaffilmark{1}}

\altaffiltext{1}{Department of Physics, Technion -- Israel
Institute of Technology, Haifa 32000, Israel;
soker@physics.technion.ac.il}

\begin{abstract}
I propose that some irregular `messy' planetary nebulae owe their
morphologies to triple-stellar evolution where tight binary
systems evolve inside and/or on the outskirts the envelope of
asymptotic giant branch (AGB) stars. {{{{ In some cases the tight
binary system can survive, in other it is destroyed. }}}} The
tight binary system might breakup with one star leaving the
system. In an alternative evolution, one of the stars of the
brook-up tight binary system falls toward the AGB envelope with
low specific angular momentum, and drowns in the envelope. In a
different type of destruction process the drag inside the AGB
envelope causes the tight binary system to merge. This releases
gravitational energy within the AGB envelope, leading to a very
asymmetrical envelope ejection, with an irregular and `messy'
planetary nebula as a descendant. The evolution of the
triple-stellar system can be in a full common envelope evolution
(CEE) or in a grazing envelope evolution (GEE). Both before and
after destruction {{{{ (if take place) }}}} the system might lunch
pairs of opposite jets. One pronounced signature of triple-stellar
evolution might be a large departure from axisymmetrical
morphology of the descendant planetary nebula. I estimate that
about one in eight non-spherical PNe is shaped by one of these
triple-stellar evolutionary routes.
\end{abstract}


% ==========================================================
\section{INTRODUCTION}
\label{sec:intro}
% ==========================================================

Planetary nebulae (PNe) are believed by many to acquire their
non-spherical morphologies through the interaction of their
progenitor with stellar or sub-stellar companions (e.g, see
summary by \citealt{DeMarcoSoker2011}). Single stars cannot
account for many of the observed properties of elliptical and
bipolar PNe \citep{Soker1998}, and single stars cannot maintain a
fast rotation over the evolution along the AGB (e.g.,
\citealt{SokerHarpaz1992, NordhausBlackman2006,
GarciaSeguraetal2014}). Following earlier studies, (e.g.,
\citealt{BondLivio1990, Bond2000}), the binary shaping paradigm
has gain a critical support in recent years (see reviews by
\citealt{Zijlstra2015} and \citealt{DeMarco2015}), in particular
due to observations of many close binary systems of elliptical and
bipolar PNe and a careful analysis of their morphologies, e.g.,
\cite{Akrasetal2015}, \cite{Alleretal2015a, Alleretal2015b},
\cite{Boffin2015}, \cite{Corradietal2015}, \cite{DeMarcoetal2015},
\cite{Douchinetal2015}, \cite{Fangetal2015},
\cite{Hillwigetal2015}, \cite{Jones2015}, \cite{Jonesetal2015},
\cite{Manicketal2015}, \cite{Martinezetal2015},
\cite{Miszalskietal2015}, \cite{Mocniketal2015},
\cite{Montezetal2015}, limiting the list to papers from 2015.
These close binary systems went through a common envelope
evolution (CEE). Some other non-spherical nebulae around AGB and
post-AGB stars might have been shaped by wider binary companions
that did not go through a CEE (e.g., \citealt{Lagadecetal2011,
Bujarrabaletal2013, VanWinckeletal2014, Decinetal2015}).

Despite much progress (e.g., \citealt{Ivanovaetal2013}), the CEE
of these systems is not well understood. Numerical simulations
cannot eject the envelope in a consistent and persistent manner.
Instead, a circumbinary flattened envelope is typically formed,
the envelope is not entirely ejected, and spiraling-in ceases too
early (e.g., \citealt{SandquistTaam1998, Lombardi2006,
DeMarco2011, Passy2011, Passyetal2012, RickerTaam2012}). In light
of these difficulties (e.g., \citealt{Soker2013}) it has been
suggested that in many cases jets launched by the more compact
companion, including main sequence secondary stars, facilitate
envelope ejection in a CEE (e.g., \citealt{Soker2013, Soker2014}).

If the jets launched by the secondary star remove the entire
envelope outside the secondary orbit as it spirals in, the
formation of a CE is prevented. Instead, a grazing envelope
evolution (GEE) takes place, where the binary system might be
considered to evolve in a state of "just entering a CE phase"
\citep{Soker2015}. The companion grazes the envelope of the giant
star, and both the orbital separation and the giant radius shrink
simultaneously. Tidal interaction that leads to orbit shrinkage
makes the GEE an alternative to the CEE in cases where jets
efficiently remove the envelope. When the companion is massive
enough to bring the giant envelope to synchronization, the orbit
during the GEE does not shrink much, and might even increase.
\cite{Gorlovaetal2012} found that the companion to the post-AGB
star BD$+46^\circ442$ launches jets. The orbital period of this
binary system is $140.77~$days. \cite{Gorlovaetal2015} reported
the detection of a collimated outflow from the companion of
IRAS~19135+3937, a post-AGB binary system with an orbital period
of 127 days. Both these systems are highly compatible with the
expectation from the GEE when the companion manages to maintain
the AGB star in synchronization with the orbital period. I suggest
that these systems evolved indeed through a synchronized GEE.

The multitude of processes that are involved in the CEE and GEE
phases, such as mass lose through the $L_2$ Lagrangian point and
jets launching with an without precession, and the large parameter
space of binary system properties, ensure that each PN is `unique'
in its morphology \citep{Soker2002}. As well, the multitude of
processes place PNe on the crossroad of many astrophysical
objects, e.g., novae, symbiotic binaries, and massive binary
systems such as $\eta$ Carinae, and of many generic astrophysical
evolutionary phases, e.g., CEE, jet-inflated bubbles, and stellar
merger \citep{DeMarco2015}. PNe are directly connected also to
type Ia supernovae \citep{TsebrenkoSoker2015} and to intermediate
luminosity optical transients (ILOTs; \citealt{SokerKashi2012}).
These processes and objects involve strongly interacting binary
systems.

On top of the large parameter space of binary system properties, a
tertiary star can be added. In previous papers mainly wide
tertiary stars were mentioned. Namely, the secondary and tertiary
stars orbit the AGB progenitor of the PN, and not each other. A
wide stellar companion, to a central single star or a central
binary system, at orbital separations of $a_3 \approx 10-{\rm
several}\times 10^3 \AU$ can cause the PN to posses departure from
axisymmetry and to have an equatorial spiral pattern
\citep{Soker1994}. \cite{Sokeretal1992} proposed that the
departure of the PN NGC~3242 from axisymmetry was caused by a
tertiary star of a mass of $M_3 \approx 1 M_\odot$ at an orbital
separation of $a_3 \approx 4000 \AU$. In addition, a very wide
tertiary star, $a_3 \ga 10^3 \AU$ might form a small bubble inside
the nebula \citep{Soker1996}.

Engulfed binary systems were discussed in the past as well.
\cite{Exteretal2010} proposed that the AGB stellar progenitor of
the PN SuWt~2 engulfed a tight binary system of A stars. They (see
also \citealt{Bondetal2002}) proposed that PNe with a high density
ring in the equatorial plane can result from triple-stellar
evolution. In that scenario the tight binary is engulfed by the
primary star, and survives most or all of the CEE. It might merge
later on as the stars of the binary system evolve
\citep{Bondetal2002}.

In the present paper I add more to this rich variety of binary and
triple evolutionary tracks by considering new evolutionary routes
were instead of a single secondary star, a tight binary system is
swallowed by the AGB or red giant branch (RGB) star. The typical
initial orbital period of the tight binary system is $P_{\rm ZAMS,
23} \approx$day$-$month, while the orbital period of the triple
system (the tight binary and the primary star around their mutual
center of mass) is $P_{ZAMS,123} \approx 1-10 \yr$. Such systems
are known to exist (e.g., \citealt{Tokovininetal2006}). This
preliminary study sets down the foundations for further studies,
including three-dimensional hydrodynamical simulations of
triple-CEE and triple-GEE. In section \ref{sec:setting} I describe
the triple-stellar systems that are considered here. In section
\ref{sec:outcomes} I describe the different possible fates of such
triple systems, and the general departure from symmetry of the
descendant PNe. I summarize in section \ref{sec:Summary}.

% ==========================================================
\section{SETTING THE PROBLEM}
 \label{sec:setting}
% ==========================================================
% ==========================
\subsection{Expectation}
 \label{subsec:expectation}
% ==========================
Let us estimate the fraction of AGB stellar progenitors of PNe
that interact with a tight binary system, rather than with a
single secondary star. Consider main sequence stars that are
expect to evolve along the upper AGB, hence having a zero age main
sequence (ZAMS) mass of $M_{\rm ZAMS,1} \simeq 1-8 M_\odot$.
Consider only those primary stars that have a main sequence
stellar companion with a mass of $M_2 < M_{\rm ZAMS,1}$, with an
initial semi-major axis $a_{\rm ZAMS,123}$. The question is the
following. What fraction of these systems have a tertiary star
with a mass of $M_3<M_2$ that orbit the secondary star in a tight
orbit, $a_{\rm ZAMS,23} <a_{\rm ZAMS,123}$? I call this fraction
$\eta_{1,23}$.

In their review article \cite{DucheneKraus2013} give for stars
with ZAMS mass in the range $M_{\rm ZAMS,1} = 1.5-5 M_\odot$ a
multiplicity frequency of MF$>50 \%$ and a companion frequency of
CF$=100 \pm 10 \%$ {{{{ (their Table 1 and figure 1) }}}}. {{{{
For solar type stars the fraction of systems with $n \le 6$ stars
goes approximately as $N(n) \approx k {2.5}^{-n}$, and $\approx 25
\%$ of systems of solar type stars have $n \ge 3$
(\citealt{DucheneKraus2013}, section 3.1.5). For the mass range
$1.5-5 M_\odot$ that is more relevant for Galactic PNe, }}}} I
take the fraction of systems with $n$ stars to go as $N(n) = k
{x}^{-n}$ as well. {{{{ I limit the systems for up to 6 stars. I
find that to account for a MF$>50 \%$ and a companion frequency of
CF$=100$ the value is $x \approx 1.9$ (taking $x=2$ that gives
CF$=90$ does not change the results much).  The fractions of
systems with $n$ stars, $N(n) \propto 1.9^{-n}$, are as follows.
Single star systems $N(1)=0.49$, binary systems $N(2)=0.25$,
triple systems, $N(3)=0.13$, and then $N(4)=0.07$, $N(5)=0.04$,
and $N(6)=0.02$. These values give MF$=51 \%$ and CF$=98 \%$. }}}}

Let us first consider only binary and triple systems. In the
triple-stellar systems the tertiary star can orbit the secondary
star, the primary star, or both. {{{{ We now consider the first
possibility which can lead to a case where an AGB star swallows a
tight binary system of two main sequence stars. I crudely assume
that in about third of all triple-stellar systems the initial more
massive star is at a larger orbital separation from the two
lighter ones. This is a fraction of $\simeq N(3)/3 =0.043$ of all
systems. }}}} The rest, $\approx 0.09$ of all systems, are
triple-stellar systems that are not the triple systems consider
now. The fraction of the systems considered now to all cases with
binary interaction under these assumptions is
\begin{equation}
 \eta_{1,23} = \frac{\rm tight~ (M_2,M_3)~sytems}{\rm all ~ multiple~ systems}
 \approx \frac {N(3)/3}{N(2)+ N(3)} \approx  \frac {0.043}{0.25 + 0.13} = 0.11 .
 \label{eq:eta123A}
\end{equation}

{{{{ Three }}}} considerations increase the fraction of PNe shaped
by triple-stellar systems relative to the value of $\eta_{1,23}$
given by equation (\ref{eq:eta123A}). Firstly, when the tertiary
star and the primary star closely orbit each other, the tertiary
might prevent the primary from formation a PN by causing a too
early envelope removal. {{{{ This reduce the denominator in
equation (\ref{eq:eta123A}). }}}} Secondly, when the higher
multiple systems are considered the probability of the AGB primary
star to interact with a binary system rather than with a single
secondary star increases. If for example, third of the quadruple,
quintuplet, {{{{ and sextuplet }}}}  stellar systems are taken to
be relevant to the present study, then
\begin{equation}
 \eta_{1,23} \approx \frac {[N(3)+N(4)+N(5)+N(6)]/3}{N(2)+ N(3)+N(4)+ N(5)+N(6)} \approx  \frac {0.26/3}{0.51} = 0.17.
 \label{eq:eta123B}
\end{equation}
{{{{ Thirdly, in some systems one of the stars of the tight binary
system might be a WD. This is the case when the initially most
massive star is in a tight orbit, and it forms a WD without
destroying the tight binary system. The initially second most
massive star is in a larger orbital separation from the tight
binary system. When it evolves on the upper AGB it might swallow
the tight binary system, now composed of a WD and a main sequence
star. }}}}

{{{{ Two considerations reduce the fraction of PNe shaped by
triple-stellar systems relative to the value of $\eta_{1,23}$
given by equations (\ref{eq:eta123A}) and (\ref{eq:eta123B}).
Firstly, some PNe can result from solar-like stars. These have
lower multiplicity frequency and companion frequency than stars
with initial mass in the range $1.5-5 M_\odot$
\citep{DucheneKraus2013}. Secondly, some PNe might be shaped by
planets. I assume here that single star do not form bright PNe.
But planets might shape PNe as well, adding to non-spherical PNe,
hence increasing the denominator in equations (\ref{eq:eta123A})
and (\ref{eq:eta123B}). On the other hand, one then needs to
consider AGB stars that swallow planetary systems. Namely, instead
of a tight binary system, the AGB swallows a tight planetary
system. Such systems can also shape PNe simpler to triple-stelar
systems if the planet is massive enough. }}}}

{{{{ Taking these considerations and the large uncertainties  I
estimate that about one in ten to one in six }}}} ($\approx 10 \%
- 17 \%$) of PN progenitors that evolved through a binary
interaction, actually went through an interaction with a tight
binary system. {{{{ Crudely, about one in eight }}}} PNe shaped by
binary systems was actually shaped by a tertiary system.

{{{{ In this work I consider mainly cases where the tight binary
system does not survive the CEE. It is not crucial for the general
results, but it is my expectation that }}} } in most, but not all,
cases of CEE only two stars survive as a bound binary system. One
of the two stars in the tight binary system, mostly the tertiary
star, is expected to merge either with the secondary star or the
primary core, or be ejected from the system. In a minority of
cases all three stars are expected to survive. {{{{ This
hypothesis that most tight binary systems do not survive, should
be examined by detail numerical simulations. }}}}

{{{{ One case where the tight binary system seems to have survived
the CEE is the PN SuWt~2 \citep{Exteretal2010}. }}}} A possible
example for evolution where the tight binary system did not
survived the CEE, but all stars survived and stayed bound is the
triple pulsar system PSR~J0337+1715. The pulsar is orbited by two
low mass white dwarfs having periods of 1.63 day and 327 day
\citep{Ransometal2014}. \cite{SabachSoker2015} proposed the
following scenario for this system (for an alternative scenario
see \citealt{TaurisHeuvel2014}). A tight binary system of two main
sequence stars was tidally and frictionally destroyed inside the
envelope of a massive star that formed a PN. Only later in the
evolution the central massive ONeMg white dwarf experienced an
accretion-induced collapse and formed the neutron star. That
scenario includes a new ingredient of a binary system that breaks
up inside a CE. As well, the scenario employs an efficient
envelope removal by jets launched by the main sequence stars
immersed in the giant envelope, and the GEE.

% ==========================
\subsection{Energy considerations}
 \label{subsec:energy}
% ==========================

Two types of energy can be associated with the tight binary
system. The first is the energy that is need to breakup the tight
binary system, the breakup energy
\begin{equation}
  E_{\rm b, 23} = \frac {1}{2} \frac {G M_2 M_3}{a_{\rm ZAMS,23}}.
 \label{eq:Eb23}
\end{equation}
The second one is the energy the binary system releases if it
merges, the merger energy
\begin{equation}
  E_{\rm m, 23} \approx \frac {1}{2} \frac {G M_2 M_3}{R_2},
 \label{eq:Em23}
\end{equation}
where $R_2$ is the radius of the secondary star. The amount of
energy released depends on the response of the secondary star to
the accretion of the tertiary star, so the value given above is an
estimate.

More constructive is to define dimensionless quantities. Let
$a_{\rm f, CE}$ be the final orbital separation of the secondary
star from the core of the primary star according to the energy
prescription of the CEE (e.g., \citealt{Webbink1984,
TaurisDewi2004, Ivanovaetal2013}) that states
\begin{equation}
  E_{\rm env} = \frac {1}{2} \alpha_{\rm CE} \frac {G M_{\rm core} M_2} {a_{\rm f,
  CE}},
 \label{eq:af}
\end{equation}
where $E_{\rm env}$ is the binding energy of the giant envelope,
$\alpha_{\rm CE}$ is the CE-$\alpha$ parameter, and $M_{\rm core}$
is the mass of the giant core. The two dimensionless ratios are
\begin{equation}
  \xi_b \equiv \frac { E_{\rm b, 23}}{E_{\rm env}} \approx 0.5
   \left( \frac{M_3}{0.5 M_{\rm  core}}\right)
  \left( \frac{2 a_{\rm f, CE}} {a_{\rm ZAMS,23}}  \right)
  \left( \frac{\alpha_{\rm CE}}{0.5} \right)^{-1},
 \label{eq:xib}
\end{equation}
and
\begin{equation}
  \xi_m \equiv \frac { E_{\rm m, 23}}{E_{\rm env}} \approx 5
  \left( \frac{M_3}{0.5 M_{\rm core}} \right)
  \left( \frac{a_{\rm f, CE}}  {5 R_2} \right)
 \left( \frac{\alpha_{\rm CE}}{0.5} \right)^{-1}.
 \label{eq:xim}
\end{equation}
Typical values of $M_{\rm core}\simeq 0.6 M_\odot$, ${a_{\rm f,
CE}} \approx 2.5 R_\odot$ (final core-secondary orbital period of
about half a day), $R_2=0.5 R_\odot$ ($M_2 \approx 0.5 M_\odot$),
${a_{\rm ZAMS,23}} \approx 5 R_\odot$, and $M_3 \approx 0.1-0.5
M_\odot$ were used for the above scaling.

Equation \ref{eq:xib} shows that the breakup of the tight binary
requires a non-negligible amount of energy. This comes on the
expense of the kinetic energy of the secondary or the tertiary
star; one star of the tight binary moves to larger orbit, or even
escape the system, and the other star dives further in.

Equation \ref{eq:xim} shows that a merger of the tight binary
system can in principle eject a large fraction of the giant
envelope, and at high speeds. The merger occurs on several times
the dynamical time of the tight-binary system, which is much
shorter than the orbital period of the triple-system. This implies
a very asymmetrical mass ejection to one side. If the two orbital
planes, of the tight binary system and of the triple system, are
parallel, the ejection process maintains the mirror symmetry about
the orbital plane. But if the two planes are inclined to each
other, no axis or plane of symmetry are expected in the mass
ejection due to the tight-binary merger. A `messy' expanding shell
is expected to be formed.

Since most PN central stars have masses in the range $M_{\rm core}
\approx 0.6-1 M_\odot$, equations \ref{eq:xib} and \ref{eq:xim}
with the initial constraints $M_3<M_2<M_{\rm ZAMS,1}$ show that on
average the role of the tight-binary system is expected to
increase with primary mass, as the mass of $M_3$ can be larger. In
the case of the tight-binary merger process even a tertiary of
$M_3 \approx 0.05-0.1 M_{\rm core}$ can play a significant role.
This includes a brown dwarf tertiary.

 The energy of the merger can be carried by jets, or a wide
outflow, launched by the accretion disk or belt formed around the
secondary star from the destroyed tertiary star. Another source of
energy, that can exist also in the case of a single secondary star
(namely, a binary system), is the accretion of giant stellar
envelope mass onto the secondary (or tertiary) star. The accretion
disk or belt can lunch jets that facilitate the removal of the
giant envelope. The energy carried by the jets can be expressed as
\citep{Soker2014}
\begin{equation}
\xi_{\rm jets} \equiv \frac { E_{\rm jets}}{E_{\rm env}} \approx
 \frac{M_{\rm acc}}{M_{\rm core}}
 \frac{a_{\rm f, CE}}  {R_2}
 \frac{1}{\alpha_{\rm CE}}
= 1
 \left( \frac{M_{\rm acc}}{0.1M_{\rm core}} \right)
 \left( \frac{a_{\rm f, CE}}{5R_2} \right)   \left( \frac{\alpha_{\rm CE}}{0.5} \right)^{-1},
\label{eq:ejet1}
\end{equation}
where $M_{\rm acc}$ is the mass accreted by the secondary (or
tertiary) star.

% ==========================================================
\section{OUTCOMES}
\label{sec:outcomes}
% ==========================================================

% =============================
\subsection{Tight binary in a wide orbit}
 \label{subsec:wide}
% =============================

% ==========
\subsubsection{Evolution}
 \label{subsubsec:EVwide}
% ==========

The first scenario for a giant star orbited by a tight binary
system evolves no CEE or GEE of a giant with a tight binary
system, but rather accretion by the tight binary system from the
dense wind of the evolved giant star. It was studied in the past
(\citealt{Soker2004} where more details can be found), and it is
updated here, as the PNe listed then might not be compatible with
new expectations.

In that study it was found that when the orbital plane of the
accreting tight binary system and the orbital plane of the triple
system (the plane of the tight binary and the giant star) are not
parallel to each other, the accreted mass onto one or two of the
tight binary system components has high specific angular momentum.
For a large fraction of triple-stellar systems accretion disks are
expected to form around one or two of the tight binary stars, and
lunch jets. The axis of jets will be almost parallel to the
orbital plane of the triple-stellar system. One jet is blown
outward relative to the wind, while the other jet passes near the
mass-losing star, and is more likely to be deflected and slowed
down by the giant wind. This flow of asymmetrical jets breaks the
mirror symmetry of the descendant PN. Namely, the two lobes or two
opposites blobs on the two sides of the equatorial plane will not
be equal. The long-period orbital motion will lead to a departure
from axisymmetry, i.e., there is no symmetry around the angular
momentum axis (the orbital motion itself does not break the mirror
symmetry about the equatorial plane).

At the end of the AGB phase when the mass-loss rate is very high,
an accretion disk might form in tight binary systems up to
$\approx ~400-800 \AU$ from the giant star. This leaves a space
for a fourth star to orbit the AGB star at a close orbit of few
AU. The closer stellar companion to the AGB star strongly
interacts with the AGB star and leads to the formation of an
elliptical or a bipolar PN. Again, the AGB star itself engulfs
only one star, if at all. As we see next, it seems that the fourth
star is required to form a PN.

% ==========
\subsubsection{The descendant PNe}
 \label{subsubsec:PNwide}
% ==========

All three stars survive. The triple-system orbit increases due to
the mass lost from the AGB stars. It can be by a factor of
$\approx 2-3$, depending on the initial and final masses.  The
orbit of the tight binary system decreases by a small amount as a
result of mass accretion.

The following PNe were listed in 2004 \citep{Soker2004} as
possessing morphologies compatible with the expectation from a
far-away tight binary system that accretes mass from the wind and
launches jets.
\newline
 \textbf{He~3-1357 (PN G331.3-12.1).} This PN has a dense ring and small lobes protrude from the nebula
\citep{Bobrowskyetal1998}. However, no single axis of symmetry,
nor a point-symmetry, and nor a plane of symmetry, can be defined
for this nebula.
\newline
 \textbf{PN NGC~6210 (PN G043.1+37.7).} This PN, as another example of a PN that was proposed to
have been shaped by a triple-stellar system \citep{Soker2004}, is
a `messy' PN, with a general elliptical structure with unequal
sides, blobs, filaments and two pairs of opposite jets protruding
from the main `messy' (irregular) shell (e.g.,
\citealt{Balick1987, Pottaschetal2009}; I will return to this PN
below).
\newline
\textbf{IC~2149 (PN G166.1+10.4)} \citep{Vazquez2002};
\textbf{M1-59 (PN G023.9-02.3)}  \citep{Manchadoetal1996,
Sahaietal2011}; \textbf{NGC~1514 (PN G165.5-15.2)}
\citep{Hajianetal1997, MuthuAnandarao2003}; \textbf{NGC~6886 (PN G
060.1-07.7)} \citep{Sahaietal2011}. These four PNe were also in
that list or PNe shaped by triple or quadruple star systems.

However, in the years since 2004 a picture has emerged that for a
bright PN to be formed the mass loss from the AGB star must be
enhance by a strong binary interaction \citep{DeMarcoetal2004,
SokerSubag2005, MoeDeMarco2006}. As such, it is less likely that
the above listed PNe are compatible with a far-away triple system
alone. They might be more compatible with the engulfment of a
tight binary system (as suggested below for NGC~6210). We
therefore need to consider one of two scenarios.

In the first scenario the tight binary system is close to the AGB
star, and accretes mass from the AGB envelope via a Roche lobe
over flow rather than from a wind. The binary system is massive
enough to maintain the triple system in synchronization and
against the Darwin instability, and the triple-stellar system
survives the entire evolution with no CEE of GEE phases. The
outcome is a point symmetric bipolar PNe with a triple star system
inside: a tight binary system orbiting the remnant of the AGB core
at $\approx 1 \AU$. In the case that the two orbital planes are
parallel, the morphology of the descendant PN will be more or less
as expected in binary evolution, but possibly with a wide and
pronounced jets' precession. When the two orbital planes are
inclined to each other, a point-symmetric PN with complicated
structure that is hard to predict will emerge.

\begin{figure}[!t]
\centering
\includegraphics[trim= 0.0cm 0.0cm 0.0cm 0.0cm,clip=true,width=0.7\textwidth]{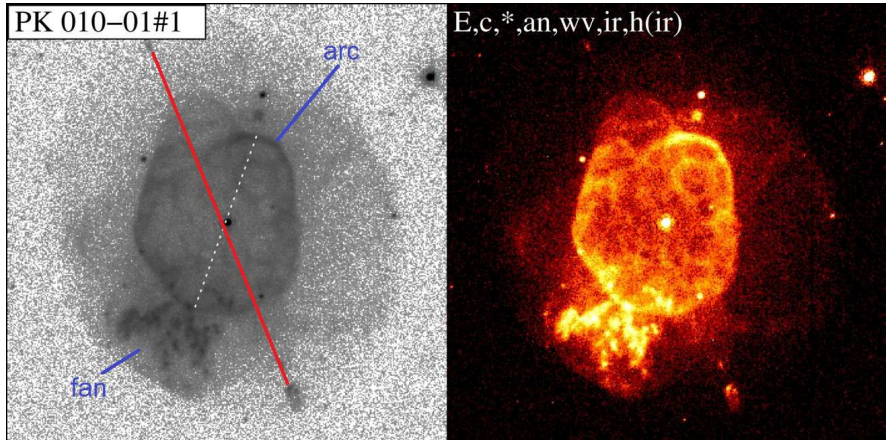}     %trim=l b r t
\caption{An image of NGC~6578 from \cite{Sahaietal2011}. This
morphology might be compatible with a tight binary system in a
wide orbit. According to the proposed scenario, both the closer
fourth star and the far-away tight binary systems, each launched a
pair of jets. The two axes of the pairs are inclined to each
other. One axis marked by \cite{Sahaietal2011} in dashed-white,
and the other added here as a red solid line. }
 \label{fig:NGC6578}
\end{figure}

In the second scenario there is a quadruple-stellar system. The
fourth star is closer to the AGB star than the far-away tight
binary system. For the tight binary system to make a mark on the
descendant PN, it should be closer than $\approx 100 \AU$. I
expect the tight binary system to add an opposite pair of bullets
or two jet-inflated small bubbles. As well, the motion around the
center of mass of the quadruple systems leads to departure from
axisymmetry. Since we are now considering quadruple systems, such
PNe are very rare. It is hard to define the exact characteristics
of the descendant PNe, and more so to find one. Let me give one
example that might be compatible with a quadruple system with a
wide tight binary system
\newline
\textbf{NGC~6578 (PN G010.8-01.8)} \citep{Palenetal2002,
Sahaietal2011}. This is an elliptical PN containing parts with not
well define symmetry  \citep{Sahaietal2011}. The PN is presented
in Fig. \ref{fig:NGC6578}. It has two pairs of opposite unequal
structures. One pair has a small lobe on each side of the center,
but they are unequal in size and shape. Outside there are two
opposite bullets, marked with the red line on the figure. The
other pair has dense structures, one is an arc and the other is a
`hand fan' of filaments. The two axes are inclined to each other.
It is hard to tell at this stage which axes is due to the closer
fourth star, and which is from the far-away tight binary system.
% FFFFFFFFFFFFFFFFFFFFFFFFFFFFFFFFFFFFFFFFFFFFFFFFFFFFFFFFFFFFFFFFFF
% FFFFFFFFFFFFFFFFFFFFFFFFFFFFFFFFFFFFFFFFFFFFFFFFFFFFFFFFFFFFFFFFFF

% =============================
\subsection{Ejecting a star}
 \label{subsec:largeorbit}
% =============================
% ==========
\subsubsection{Evolution}
 \label{subsubsec:EVlargeorbit}
% ==========

In the interaction of three point masses one possible outcome is
that the lowest mass object is ejected from the system, and the
two other bodies form a bound binary system. The case where the
primary star has an envelope of the size of the orbital orbit and
the tight binary system breaks up inside the giant envelope has
not been studied numerically. The breakup of the binary system was
discussed somewhat speculatively by \cite{SabachSoker2015}, as
summarized in section \ref{subsec:expectation} here.
\cite{SabachSoker2015} were interested in the case when the two
orbital planes are parallel to each other. Here we refer to
inclined planes as well.

The initial system is composed of a tight binary system with an
orbital separation of $a_{\rm ZAMS,23} \approx {\rm several}
\times 10 R_\odot$, that orbit a primary star of mass $M_{\rm
ZAMS,1} \simeq 2-6 M_\odot$, at a triple system orbital separation
of $a_{\rm ZAMS,123} \approx {\rm several} \times a_{\rm ZAMS,23}
\approx 1-5 \AU$. The range of the primary mass is dictated by the
condition that its maximum radius on the RGB will be much smaller
than its maximum radius on the AGB when engulfment supposes to
take place. The ratio of the two orbital separations ensures
dynamical stability  (e.g. \citealt{MardlingAarseth2001}) before
engulfment occurs, and a marginal stability when engulfment starts
after the triple system orbit has decreased.

At the upper AGB the primary star expands. Due to tidal forces the
tight binary system spirals-in toward the primary star. Now the
ratio of the triple system orbital separation to the tight binary
orbital separation decreases, and the system might become
dynamical unstable. If the tight binary system breaks-up before
engulfment occurs, one star, most likely the lowest mass one, is
ejected from the system, and we enter a regime of binary
interaction.

In the case that the tight binary system does not breakup before
engulfment, the two stars start to orbit inside the giant
envelope. In these systems the orbital separation of the tight
binary system is large, of the order of magnitude of the density
scale hight in the giant envelope. Hence, initially only the star
closer to the primary is inside the envelope. This star might
accrete mass, forms an accretion disk, and lunches two opposite
jets. At the same time the other star is tens of $R_\odot$ outside
the giant photosphere. In the scenario discussed in this
subsection, the jets remove the envelope outside the binary
system. Namely, the system is in the GEE phase. Each of the two
stars in the tight binary system spends only part of the orbit
inside the giant envelope. The jets are expected to exist during
part of the orbital period of the tight binary system, only when
the accreting star(s) is(are) inside the envelope.

Only the star inside the envelope suffers a strong gravitational
drag. The other star feels strong tidal forces. This situation
changes after about a half orbital period, and so on. This
differential gravitational drag on the two stars and the tidal
forces acting on the star that is outside the envelope, might
change the outcome in two ways, as speculated by
\cite{SabachSoker2015}. Firstly,  instead of being ejected from
the system, the star that was supposed to escape might end at a
large, but bound and stable, orbit. Secondly, when the masses
ration $q_{32} = M_3 / M_2$ is not much below 1, then in some
cases the star $M_2>M_3$ might end up at a larger orbital
separation.

The break-up is expected to take place when the binary system is
in the outer parts of the AGB star, in a triple system orbital
separation of $a_{123} \ga 100 R_\odot$. After the ejection of one
star from the envelope, the two remaining stars of the inner
binary system, either the primary with the secondary or the
primary with the tertiary,  continue in a binary interaction. The
inner binary system can continue the GEE, hence the two stars
survive, or enter a CEE. As in binary interaction evolution, the
CEE can end with a close binary system, or in a merger.

One of the four following types of a bound system can be found at
the center of the descendant PN {{{{ in cases where the tight
binary system breaks apart. }}} }(1) A single star. This is the
case if one of the tight binary stars was ejected and the other
merged with the core. (2) A close binary system. This is the case
if one of the tight binary stars was ejected and the other
survived the CEE or the GEE phase. (3) A wide and eccentric binary
system. This is the case if one of the tight binary stars was
thrown to a large eccentric orbit, and the other merged with the
core. (4) A triple system of a wide and eccentric star, and a much
closer binary system. This is the case if one of the tight binary
stars was thrown to a large eccentric orbit, and the other
survived the CEE or the GEE phase. {{{{ If the tight binary system
survives the CEE, the central system will be of a post-AGB star
orbited by a tight binary system \citep{Exteretal2010}. }}}}

% ==========
\subsubsection{The descendant PNe}
 \label{subsubsec:PNlargeorbit}
% ==========

The descendent nebula possesses a structure of two non-spherical
structures resulting from two mass loss phases with a relative
displacement between them. The two intensive mass loss phases are
before and after the breakup of the tight binary system. Before
the breakup the mass loss occurs around the center of mass of the
triple systems. Jets are formed as the tight binary system
accretes from the AGB envelope via a Roche lobe overflow. After
the ejection of one of the stars of the tight binary system, the
mass loss continues around the center of mass of the remaining
binary system. This system moves to the opposite direction of the
ejected star. Therefore, the nebular part formed after the breakup
has a general velocity relative to the outer nebula. This
descendant PN has a displacement from axisymmetry. If the two
orbital planes are not parallel to each other, the system will not
have even a mirror symmetry about any orbital plane. {{{{ In cases
where the tight binary system survives, only the first of these
two phases will take place, and the descendent PNe will be less
complicated, although the departure from sphericity can be very
large, e.g., having an equatorial ring \citep{Exteretal2010}. }}}}

 Let us consider one example out of a very large parameter space.
Say at breakup the binary system is at $a_{123}=1 \AU$. The masses
are $M_1=3M_\odot$, $M_2=0.5 M_\odot$, and $M_3=0.3 M_\odot$. Let
us take the tertiary star to become unbound and ejected from the
system with a terminal velocity of a fraction $\chi$ of the escape
speed from the system, $v_{\rm ej,3} \approx 40 (\chi/0.5) \km
\s^{-1}$. From momentum conservation the remaining binary system
moves with respect to the center of mass  at $v_{12} \approx  3
(\chi/0.5) \km \s^{-1}$. For a nebula expanding at $\approx 10-20
\km \s^{-1}$ this leads to an observable departure from
axisymmetry.

\begin{figure}[!t]
\centering
\includegraphics[trim= 0.0cm 0.0cm 0.0cm 0.0cm,clip=true,width=0.7\textwidth]{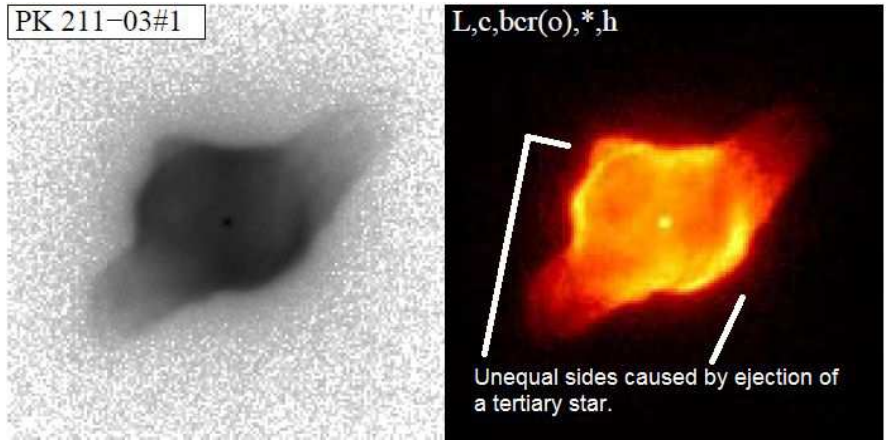}     %trim=l b r t
\caption{An image of M~1-6 from \cite{Sahaietal2011}. I suggest
that the departure from both mirror-symmetry and from axisymmetry
might have been caused by the ejection of the tertiary star from
the tight binary system. Despite these deviations, this PN is not
as `messy' as expected when the tight binary stars merge.  }
 \label{fig:M16}
\end{figure}

The PNe are not expected to be too `messy', but might posses no
symmetry axis and no plane of symmetry. Possible structures are
like those of the following PNe. (One should be aware that some
morphological features can be accounted for also in binary
interaction, with no need for triple-stellar evolution.)
 \newline
 \textbf{IC~2553 (PN G285.4-05.3).} It has a point-symmetric structure, but with
 a displacement of different structures \citep{Corradietal2000}. It is possible that a
 triple-stellar system caused this structure.
 \newline
 \textbf{M~1-26 (PN G358.9-00.7).} On top of the point-symmetric structure this PN
 possesses departure from axisymmetry along each line that connect two opposite features
 \citep{SahaiTrauger1998}. I suggest here that a triple-stellar system shaped this
 PN. An unbound low mass main sequence star might exist near the center.
  \newline
 \textbf{M~1-6 (PN G211.2-03.5).} This PN has two lobes, and a smaller and denser
 inner region \citep{Sahaietal2011, Otsukaetal2014}. Its image is presented in Fig. \ref{fig:M16} taken from \cite{Sahaietal2011}. This PN has neither axisymmetry nor mirror symmetry.
 Its structure is compatible with the expectation of a
 triple-stellar system shaping discussed in this subsection.
 \newline
 \textbf{M~3-1 (PN G242.6-11.6).} Has general departures from axisymmetry and from mirror-symmetry as those in
 M~1-6 \citep{Schwarzetal1992}.
% FFFFFFFFFFFFFFFFFFFFFFFFFFFFFFFFFFFFFFFFFFFFFFFFFFFFFFFFFFFFFFFFFF
% FFFFFFFFFFFFFFFFFFFFFFFFFFFFFFFFFFFFFFFFFFFFFFFFFFFFFFFFFFFFFFFFFF

% =============================
\subsection{Drowning star}
 \label{subsec:drowning}
% =============================
% ==========
\subsubsection{Evolution}
 \label{subsubsec:EVdrowning}
% ==========

In this evolutionary route the triple-stellar system starts more
or less as in the stellar ejection one, and the tight binary
system breaks-up as a result of dynamical instability. However,
the dynamical instability results in one star moving toward the
giant envelope, instead of to the opposite direction, and with a
low specific angular momentum. The dynamical interaction `drowns'
one of the stars of the tight binary system in the AGB envelope.
This can happen before the tight binary system touches the
envelope, or after.

We now have a triple-stellar system composed of a CEE of the AGB
star and the drowning star of the tight binary system, with a
third star orbiting at a distance of $a_{\rm out} \approx 1 \AU$.
There are four types of possible outcomes, depending mainly on the
relative masses of the three stars. (1) If the two stars of the
tight binary system are relatively massive, then the drowning star
manages to eject the AGB envelope. The outcome is a triple-stellar
system of a star orbiting the core-remnant at $a_{\rm in} \approx
1-10 R_\odot$, and the other one orbiting the central binary
system at $a_{\rm out} \approx 1 \AU$. (2) If the drowning star is
not massive enough to eject the AGB envelope, it merges with the
AGB core. The other star can then survive, either with no more
interaction with the AGB envelope, or experience a GEE phase. The
surviving star ends at a large orbital separation of $\approx 1-3
\AU$, depending on how much mass was ejected from the AGB star
during the merger process of the core with the drowning star. (3)
The drowning star does not eject enough AGB envelope mass, and the
other star enters a CEE phase. This star survives and ends like in
a regular binary interaction. (4) The drowning star does not eject
enough AGB envelope mass, and the other star enters a CEE phase.
This star is not massive enough, and it merges with the core. The
outcome is a single star system.

% ==========
\subsubsection{The descendant PNe}
 \label{subsubsec:PNdrowning}
% ==========

The main difference in the morphologies of the descendant PNe in
the present evolutionary route to those described in sections
\ref{subsec:largeorbit} and \ref{subsec:tightmerger}, is that no
large, or not at all, departure from axisymmetry, or more
generally no departure from point-symmetry, is expected. Crudely,
the evolution can be describes as composed of three intensive mass
loss episodes. ($i$) The tight binary system accretes mass from
the AGB envelope via a Roche lobe overflow. It is expected to
launch jets and form a bipolar structure. ($ii$) A CEE of the
drowning star inside the AGB envelope. Mass loss with
concentration toward the equatorial plane might take place.
($iii$) A CEE, or a GEE, or a Roche lobe overflow interaction of
the other star with the AGB star takes place. Another bipolar
structure and/or another equatorial mass loss event are expected.
If the initial plane of the tight binary system and the
triple-stellar system are not parallel to each other, the two
pairs of lobes and/or the two equatorial mass ejections will not
be parallel to each other. A large scale point-symmetric PN is
expected.

\begin{figure}[!t]
\centering
\includegraphics[trim= 0.0cm 0.0cm 0.0cm 0.0cm,clip=true,width=0.7\textwidth]{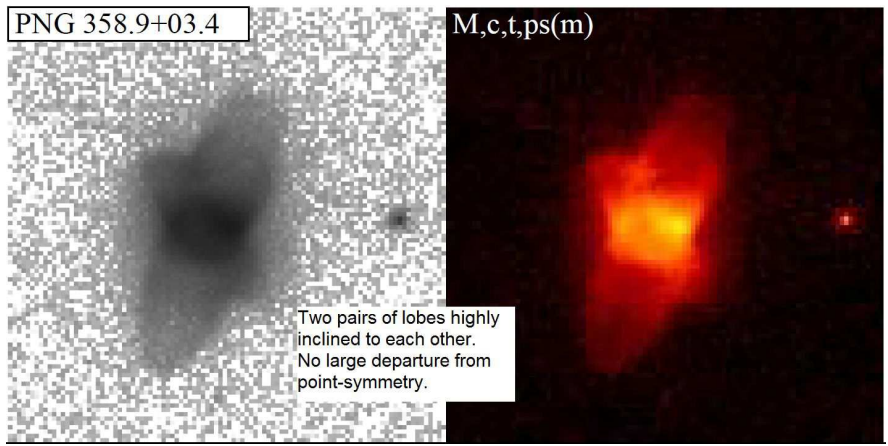}     %trim=l b r t
\caption{An image of PN~G358.9+03.4 from \cite{Sahaietal2011}. It
presents a PN with two pairs of lobes highly inclined to each
other and no large scale departure from point-symmetry. It might
be compatible with the evolutionary route of a drowning star.
 }
 \label{fig:PNG3589}
\end{figure}

The following PNe have morphologies that might be compatible with
the expectation of this evolution.
\newline
 \textbf{IPHAS~PN-1} This PN contains two pairs of lobes, one inside the other, and highly inclined to each other \citep{Mampasoetal2006}.
There is no large scale departure from point-symmetry
 \newline
 \textbf{PN~G358.9+03.4} \citep{Sahaietal2011}. This PN contains two pairs
of lobes highly inclined to each other, and one pair is larger
than the other. As evident from Fig.  \ref{fig:PNG3589} no large
departure from point-symmetry is observed.
 \newline
 \textbf{Frosty Leo (IRAS~09371+1212)} \citep{CastroCarrizoetal2005}.
This is a proto-PN, that like PN~G358.9+03.4 contains two pairs of
lobes highly inclined to each other, and one pair is larger than
the other. The small pair of lobes is actually a complicated
structure, where more small lobes are observed. This complicated
structure might come from a triple-stellar evolution.
\newline
 \textbf{NGC~6302 (PN~G349.5+01.0)} (e.g.,
\citealt{Meaburnetal2005, Szyszkaetal2011}). It has a well define
bipolar structure, but with another pair of lobes protruding from
the main bipolar structure. The axis of the protruding pair is
inclined to that of the bipolar nebula, and the two lobes of the
protruding pairs are not of equal size. \cite{SokerKashi2012}
proposed that the bipolar structure of NGC~6302 was formed in an
ILOT event.
% FFFFFFFFFFFFFFFFFFFFFFFFFFFFFFFFFFFFFFFFFFFFFFFFFFFFFFFFFFFFFFFFFF
% FFFFFFFFFFFFFFFFFFFFFFFFFFFFFFFFFFFFFFFFFFFFFFFFFFFFFFFFFFFFFFFFFF

% =============================
\subsection{Tight binary merger}
 \label{subsec:tightmerger}
% =============================
% ==========
\subsubsection{Evolution}
 \label{subsubsec:EVtightmerger}
% ==========

In this case the two stars enter a CEE or a GEE phase. The
gravitational drag on the two stars of the tight binary system is
more influential than the three-body dynamical instability, and
the two stars merge. Accretion of mass by one or two of the tight
binary stars also decreases the orbital separation. Most likely
they merge well inside the envelope. As evident from equation
(\ref{eq:xim}), the energy released in the merger process is
large. A large fraction of one side of the envelope can be ejected
at high velocities, and a binary system survives: the primary and
the secondary with the mass it accreted from the destroyed
tertiary star (only a fraction of the destroyed tertiary star is
accreted by the secondary star).

Let us consider one example from the large parameter space. Say a
secondary of mass and radius of $M_2=0.5 M_\odot$ and $R_2=0.5
R_\odot$, respectively, accretes $M_{\rm acc} = 0.1 M_\odot$ of
the destroyed tertiary star. By equation (\ref{eq:Em23}) the
released energy is $E_{\rm m, 23} \approx 2 \times 10^{47} \erg$.
The accretion process lasts for about the viscosity time of the
accretion disk. By equation (2) from \cite{KashiSoker2015} the
viscosity time is about two weeks. If this energy is channelled to
eject, say, $M_{\rm env-ej} = 0.5 M_\odot$ of the envelope mass,
the initial velocity of the ejected envelope mass is $v_{\rm
env-ej} \approx 200 \km \s^{-1}$. This is the escape speed from
$\approx 10 R_\odot$ from the core. So if the merger occurs at a
radius of $\ga 20 R_\odot$, the ejected envelope chunk is ejected
at a relatively high speed.

The fast ejected envelope mass can collide with previously ejected
mass, and kinetic energy is channelled to thermal energy and
radiation. An energy of $\approx 10^{47} \erg$ can be radiated in
several weeks. These are close to the radiated energy and time
scale of the ILOT of the AGB star NGC~300~OT2008-1 (NGC~300OT;
\citealt{Monard2008, Bond2009, Berger2009}). So one of the outcome
of this merger of the tight binary system can be an ILOT event in an AGB star.

The large released energy facilitates the envelope removal, and
the final orbital separation of the binary system, the core and
secondary star, can be relatively large. As a large mass of the
envelope is ejected with high momentum to on side, the surviving
binary system composed of the giant and the secondary star moves
to the other. In the example above, let us take the terminal
velocity of the one-sided ejected mass $M_{\rm env-ej} =0.5
M_\odot$ to be $100 \km \s^{-1}$. Let the rest of the primary mass
be $3.5M_\odot$, which together with the secondary is $4 M_\odot$.
Although the surviving system is eight times as a massive as the
ejected mass, its velocity relative to the the center of mass is
lower than $100/8=12.5 \km \s^{-1}$. This is because the ejected
envelope mass is not a point mass, but rather it spreads over a
large solid angle. Overall, the surviving binary system might move
at $\approx 5-10 \km \s^{-1}$ relative to the center of mass. The
ejection of a large envelope fraction to one side, and the motion
of the surviving system, that continues to lose mass, to the other
side results in a `messy' (irregular) PN.

% ==========
\subsubsection{The descendant PNe}
 \label{subsubsec:PNtightmerger}
% ==========

The descendant nebula has no axisymmetry, and in some cases not
even a mirror symmetry about the orbital plane. A binary system
survives, but it is displaced from the nebular center, if a center
can be defined at all for the expected `messy' PN. It is expected
to have one side much brighter than the other. The bright massive
side can be larger or smaller than the faint size, depending on
later evolution, e.g., acceleration by a fast wind.

\begin{figure}[!t]
\centering
\includegraphics[trim= 0.0cm 0.0cm 0.0cm 0.0cm,clip=true,width=0.35\textwidth]{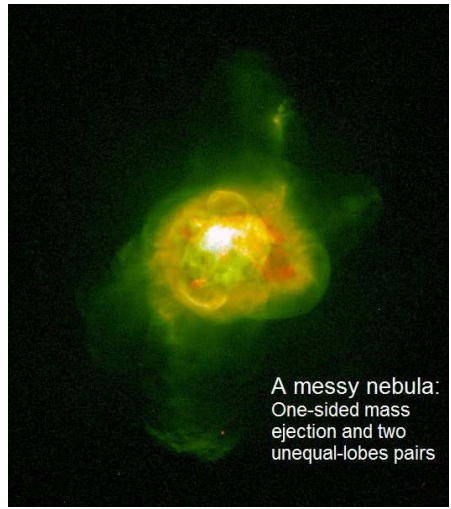}     %trim=l b r t
\caption{An image of NGC~6210 (from HST site), presenting a
`messy' PN that is compatible with the expectation from a tight
binary system that launched jets and merged. Jet launching can
occur after the merger process as well.
 }
 \label{fig:NGC6210}
\end{figure}

\begin{figure}[!t]
\centering
\includegraphics[trim= 0.0cm 0.0cm 0.0cm 0.0cm,clip=true,width=0.7\textwidth]{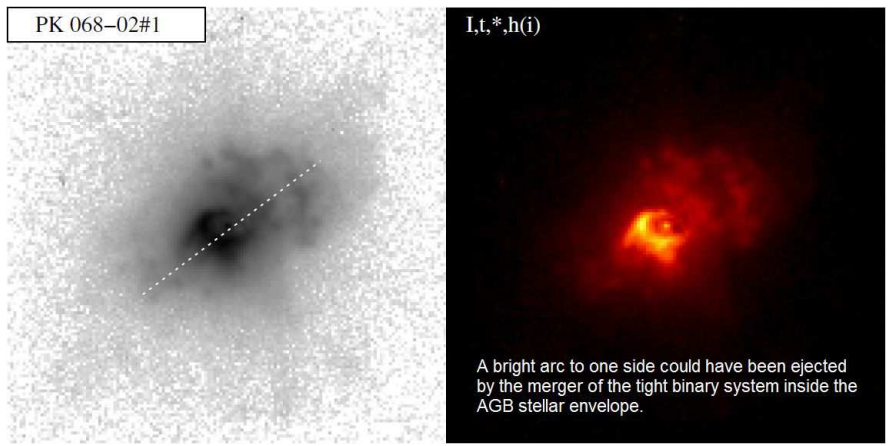}     %trim=l b r t
\caption{An image of Hen~2-459 from \cite{Sahaietal2011}. The
pronounced dipole morphology is compatible with the expectation of
mass ejected in a merger process of the tight binary system inside
the AGB envelope. }
 \label{fig:Hen2459}
\end{figure}

The following PNe have morphologies that are compatible with the
expectation of a tight binary system merger inside the progenitor
envelope.
\newline
\textbf{Sh2-71 (PN G035.9-01).} This bipolar PN is not really a
`messy' PN, but it lacks a well define symmetry axis or a symmetry
plane \citep{Mocniketal2015}. It is compatible with the
expectation from the scenario discussed here when the energy
released in the tight binary merger is not large.
\newline
\textbf{Abell~46 (PN G055.4+16.0).} In this PN the mass
distribution departs from any symmetry. The companion to the
central star has a mass of 0.15Mo \citep{Corradietal2015}.
\newline
\textbf{NGC~6210 (PN G043.1+37.7).} This was mention by
\cite{Soker2004} as descendant of a triple-stellar system with
tight binary being at a wide orbit (section \ref{subsubsec:PNwide}
above). As evident from Fig. \ref{fig:NGC6210}, this is indeed a
`messy' PN \citep{Balick1987, Pottaschetal2009}, and its
morphology is better compatible with the expectation from a tight
binary merger. The two unequal-lobes pairs might hint that the
tight binary system launched jets before it merged.
\newline
\textbf{Hen~2-11 (PN G259.1+00.9).} It posses a departure from
axisymmetry and mirror symmetry, although these deviations are not
large, and its central star is a binary system with a K-type main
sequence companion in a 0.609~d orbital period
\citep{Jonesetal2014}. The departure from pure axisymmetry and
mirror symmetry might be compatible with a process of tight binary
merger with a low mass M-type tertiary star or with a brown dwarf.
\newline
\textbf{Hen~2-459 (PN G068.3-02.7).} It has no lobes. One can
define a symmetry plane, where the mass in one side is more
extended, and in the other side there is a dense arc
\citep{Sahaietal2011}. It is presented in Fig. \ref{fig:Hen2459}.
\newline
\textbf{He~2-71 (PN G296.4-06.9)} \citep{SahaiTrauger1998};
\textbf{He~2-105 (PN G308.6-12.2)} \citep{Schwarzetal1992};
\textbf{He~2-113 (PN G321.0+03.9)} \citep{Sahaietal2000,
Sahaietal2011}. These are other PNe that lack both pure
axisymmetry and mirror symmetry, although they have one or more
pairs of lobes; some cases the two lobes are not equal.
 %% More that need further study:
 %%   He2-77; He 2-103; He 2-146
% FFFFFFFFFFFFFFFFFFFFFFFFFFFFFFFFFFFFFFFFFFFFFFFFFFFFFFFFFFFFFFFFFF
% FFFFFFFFFFFFFFFFFFFFFFFFFFFFFFFFFFFFFFFFFFFFFFFFFFFFFFFFFFFFFFFFFF
% FFFFFFFFFFFFFFFFFFFFFFFFFFFFFFFFFFFFFFFFFFFFFFFFFFFFFFFFFFFFFFFFFF
% FFFFFFFFFFFFFFFFFFFFFFFFFFFFFFFFFFFFFFFFFFFFFFFFFFFFFFFFFFFFFFFFFF

% ====================
\section{SUMMARY}
\label{sec:Summary}
% ====================

I conducted a preliminary study of the possible outcomes from the
strong interaction of PN progenitors with a close tight binary
system. I estimated that about one out of eight PNe that evolved
through binary interaction, actually experienced triple-stellar
evolution, including brown dwarfs tertiary stars (eqs.
\ref{eq:eta123A} and \ref{eq:eta123B}).
 Two types of energy can be associated with the tight binary
system, the energy that is required to break it up (eq.
\ref{eq:Eb23}), and the gravitational energy that is released if
the two stars of the tight binary system merge (eq.
\ref{eq:Em23}). These are not negligible relative to the binding
energy of the envelope. In particular, the energy released by the
merger process can be large and leads to ejection of a large
portion of the envelope in one direction and possibly with high
velocities. If part of the kinetic energy is channelled to
radiation, an ILOT event of an AGB star might take place.

I also speculated on the possible morphologies of the descendant
PNe of triple-stellar evolution. One should be aware that at this
stage the affiliation of specific PN morphologies to the different
triple-stellar evolutionary routes is highly uncertain, and is
qualitative and objective. As well, many features, like departure
from axisymmetry, might be caused also by a binary interaction
with no need for a triple-stellar system. However, binary
interactions are not expected to lead to a very large departures
from axisymmetry, and are not expected to cause any departure from
mirror symmetry.

I concentrated on four evolutionary routes. The first one is of a
tight binary system that never enters a CEE or a GEE phase, and
survives to the PN phase (section \ref{subsec:wide}). To form a
bright PN either the tight binary system must be close to the AGB
star, or a fourth star closer to the AGB star is required. In the
first case a Roche love over flow is likely to take place,
resulting in massive jets launched by the tight binary system,
that form bipolar PN. In the second case the fourth star interacts
strongly with the AGB star. Evolutionary routes with a
quadruple-stellar system are rare, and it is hard to picture the
expected morphology of the descendant PNe of such systems. I
speculated, non the less, on something like the morphology of
NGC~6578 that is presented in Fig. \ref{fig:NGC6578}.

In the second evolutionary route the tight-binary system breaks-up
as a result of the gravity of the primary star (section
\ref{subsec:largeorbit}). This occurs when the tight binary system
comes closer to the AGB star as a result of tidal forces, and a
GEE takes place. One star of the tight binary system, more likely
the tertiary star, is ejected from the system or is thrown to a
large and eccentric bound orbit. This causes the binary system
that is left, of the AGB star and the secondary star, to move in
the other direction. This leads to departure from axisymmetry of
the descendant PN. If the two orbital planes, of the tight binary
system and of the triple-stellar system, are not parallel to each
other, the PN will posses no mirror symmetry either. One of the
possible expected morphologies might be the one of M~1-6 as
presented in Fig. \ref{fig:M16}.

In the third evolutionary route the dynamical instability
breaks-up the tight binary system, and throws one star into the
AGB envelope with low specific angular momentum; this star
`drowns' in the AGB envelope (section \ref{subsec:drowning}). The
evolution can end up with one of four types of stellar systems.
The descendant PN is expected to have an axisymmetry if the
initial equatorial plane of the tight binary system is parallel to
that of the triple-stellar system, and to posses point-symmetry if
the planes are inclined to each other. No large scale departure
from axisymmetry is expected. I speculated that one possible PN
morphology is like that shown in Fig. \ref{fig:PNG3589}.

In the fourth evolutionary route (section
\ref{subsec:tightmerger}) the tight binary system merges inside
the AGB envelope. The large amount of released gravitational
energy can eject a large envelope mass in one direction. Both
before and after merging, and definitely during the merging
process, the system can launch jets. The PN has no axisymmetry and
no point symmetry. If the two orbital planes are inclined to each
other, the PN will neither posses mirror symmetry. The outcome is
a binary system with a `messy' PN. Possible PN morphologies are
like those shown in Figs. \ref{fig:NGC6210} and \ref{fig:Hen2459}.

In short, the main claim of this paper is that \emph{`messy' PNe
owe their structure to triple (or quadruple) stellar system
interaction.}

I did not discuss in details cases where the tight binary system
survived the GEE and/or the GEE phases, a scenario proposed by
\cite{Bondetal2002} and \cite{Exteretal2010}. In general the
morphologies are expected to be similar to some of the cases of
the drowning stellar scenario (section \ref{subsec:drowning}). If,
for example, the two orbital planes are parallel, then the outcome
can be as in binary evolution, but more extreme, e.g., an extreme
ring as suggested by \cite{Bondetal2002} and \cite{Exteretal2010}.
If the two planes are inclined, a point symmetric PNe might be
formed, with a pair of lobes inclined at an angle different than
$90^\circ$ to a dense equatorial ring or torus. {{{{ Namely, the
descendant PNe of surviving tight binary systems are expected to
posses axial-symmetry or point-symmetry despite the large
departure from spherical-symmetry. They will not be `messy'. }}}}

I benefitted from discussions with Amit Kashi, Erez Michaely, and
Efrat Sabach on several processes mentioned in the paper. I thank
Howard Bond, David Jones, Raghvendra Sahai, and Hans Van Winckel,
{{{{ and an anonymous referee }}}}  for valuable comments and
suggestions. I am supported by the Charles Wolfson Academic Chair.
The Planetary Nebula Image Catalogue (PNIC) compiled by Bruce
Balick was an essential tool in this study
 \newline
(http://www.astro.washington.edu/users/balick/PNIC/).

% %%%%%%%%%%%%  Refrences %%%%%%%%%%%%%%%%%%%%%%%%%%%%%%%%%%%%%%%%%%%%%%%%%%%%%%%%%%%%%%%%%%%%%%%%%%%%%%%%%%%%%

\label{lastpage}


\begin{thebibliography}{}\addcontentsline{toc}{section}{References}


\bibitem[Akras et al.(2015)]{Akrasetal2015} Akras, S., Boumis, P.,
Meaburn, J., Alikakos, J., Lopez, J. A., Goncalves, D.~R.\ 2015,
\mnras, 452, 2911

\bibitem[Aller et al.(2015a)]{Alleretal2015a} Aller, A., Miranda,
L.~F., Olgu{\'{\i}}n, L., Vazquez, R., Guillen, P.~F., Oreiro, R.,
Ulla, A., \& Solano, E.\ 2015a, \mnras, 446, 317

\bibitem[Aller et al.(2015b)]{Alleretal2015b} Aller, A., Montesinos,
B., Miranda, L.~F., Solano, E., \& Ulla, A.\ 2015b, \mnras, 448,
2822

\bibitem[Balick(1987)]{Balick1987} Balick, B.\ 1987, \aj, 94, 671

\bibitem[Berger et al.(2009)]{Berger2009} Berger, E., Soderberg, A. M., Chevalier, R. A., et al. 2009, \apj, 699, 1850

\bibitem[Bobrowsky et al.(1998)]{Bobrowskyetal1998} Bobrowsky, M.,
Sahu, K.~C., Parthasarathy, M., \& Garc{\'{\i}}a-Lario, P.\ 1998,
\nat, 392, 469

\bibitem[Boffin(2015)]{Boffin2015} Boffin, H.\ 2015, 19th European
Workshop on White Dwarfs, 493, 527

\bibitem[Bond(2000)]{Bond2000} Bond, H.~E.\ 2000, Asymmetrical
Planetary Nebulae II: From Origins to Microstructures, 199, 115

\bibitem[Bond et al.(2009)]{Bond2009} Bond, H.~E., Bedin, L.~R., Bonanos, A.~Z., Humphreys, R.~M., Monard, L.~A.~G.~B., Prieto, J.~L., \& Walter, F.~M.\ 2009, \apjl, 695, L154

\bibitem[Bond \& Livio(1990)]{BondLivio1990} Bond, H.~E., \& Livio, M.\
1990, \apj, 355, 568

\bibitem[Bond et al.(2002)]{Bondetal2002} Bond, H.~E., O'Brien,
M.~S., Sion, E.~M., Mullan, D.~J., Exter, K., Pollacco, D.~L., \&
Webbink, R.~F.\ 2002, Exotic Stars as Challenges to Evolution,
279, 239

\bibitem[Bujarrabal et al.(2013)]{Bujarrabaletal2013} Bujarrabal, V., Alcolea, J., Van
Winckel, H., Santander-Garc{\'{\i}}a, M., \& Castro-Carrizo, A.\
2013, \aap, 557, A104

\bibitem[Castro-Carrizo et al.(2005)]{CastroCarrizoetal2005} Castro-Carrizo, A., Bujarrabal,
V., S{\'a}nchez Contreras, C., Sahai, R., \& Alcolea, J.\ 2005,
\aap, 431, 979

\bibitem[Corradi et al.(2015)]{Corradietal2015} Corradi, R.~L.~M.,
Garc{\'{\i}}a-Rojas, J., Jones, D., \& Rodr{\'{\i}}guez-Gil, P.\
2015, \apj, 803, 99

\bibitem[Corradi et al.(2000)]{Corradietal2000} Corradi, R.~L.~M.,
Gon{\c c}alves, D.~R., Villaver, E., Mampaso, A., \& Perinotto,
M.\ 2000, \apj, 542, 861

\bibitem[Decin et al.(2015)]{Decinetal2015} Decin, L., Richards, A.~M.~S.,
Neufeld, D., Steffen, W., Melnick, G., \& Lombaert, R.\ 2015,
\aap, 574, A5

\bibitem[De Marco(2015)]{DeMarco2015} De Marco, O.\ 2015, in
``Physics of Evolved Stars 2015 - A conference dedicated to the
memory of Olivier Chesneau'', in press

\bibitem[De Marco et al.(2004)]{DeMarcoetal2004} De Marco, O., Bond,
H.~E., Harmer, D., \& Fleming, A.~J.\ 2004, \apjl, 602, L93

\bibitem[De Marco et al.(2015)]{DeMarcoetal2015} De Marco, O., Long,
J., Jacoby, G.~H., Hillwig, T., Kronberger, M., Howell, S.~B.,
Reindl, N., Margheim, S.\ 2015, \mnras, 448, 3587

\bibitem[De Marco et al.(2011)]{DeMarco2011} De Marco, O., Passy, J.-C., Moe, M., Herwig, F., Mac Low, M.-M., \& Paxton, B.\ 2011, \mnras, 411, 2277

\bibitem[De Marco \& Soker(2011)]{DeMarcoSoker2011} De Marco, O., \& Soker, N.\
2011, \pasp, 123, 402

\bibitem[Douchin et al.(2015)]{Douchinetal2015} Douchin, D., De Marco,
O., Frew, D.~J., Jacoby, G.~H., Jasniewicz, G., Fitzgerald, M.,
Passy, J-C., Harmer, D., Hillwig, T., \& Moe, M.\ 2015, \mnras,
448, 3132

\bibitem[Duch{\^e}ne \& Kraus(2013)]{DucheneKraus2013} Duch{\^e}ne, G., \& Kraus, A.\ 2013, \araa, 51, 269

\bibitem[Exter et al.(2010)]{Exteretal2010} Exter, K., Bond, H.~E.,
Stassun, K.~G., Smalley, B., Maxted, P.~F.~L., \& Pollacco, D. L.\
2010, \aj, 140, 1414

\bibitem[Fang et al.(2015)]{Fangetal2015} Fang, X., Guerrero, M.~A.,
Miranda, L.~F., Riera, A., Velazquez, P.~F., Raga, A. C.\ 2015,
\mnras, 452, 2445

\bibitem[Garc{\'{\i}}a-Segura et al.(2014)]{GarciaSeguraetal2014}
Garc{\'{\i}}a-Segura, G., Villaver, E., Langer, N., Yoon, S.-C.,
\& Manchado, A.\ 2014, \apj, 783, 74

\bibitem[Gorlova et al.(2012)]{Gorlovaetal2012} Gorlova, N., Van Winckel, H.,
Gielen, C., et al.\ 2012, \aap, 542, A27

\bibitem[Gorlova et al.(2015)]{Gorlovaetal2015} Gorlova, N., Van
Winckel, H., Ikonnikova, N.~P., Burlak, M.~A., Komissarova, G.~V.,
Jorissen, A., Gielen, C., Debosscher, J., \& Degroote, P.\ 2015,
\mnras, 451, 2462

\bibitem[Hajian et al.(1997)]{Hajianetal1997} Hajian, A.~R., Frank,
A., Balick, B., \& Terzian, Y.\ 1997, \apj, 477, 226

\bibitem[Hillwig et al.(2015)]{Hillwigetal2015} Hillwig, T.~C., Frew,
D.~J., Louie, M.,  De Marco, O., Bond, H.~E., Jones, D., Schaub,
S.~C.\ 2015, \aj, 150, 30

\bibitem[Ivanova et al.(2013)]{Ivanovaetal2013} Ivanova, N., Justham, S., Chen, X., et al.\ 2013, \aapr, 21, 59

\bibitem[Jones(2015)]{Jones2015} Jones, D.\ 2015, in
``Physics of Evolved Stars 2015 - A conference dedicated to the
memory of Olivier Chesneau'', in press (arXiv:1507.05447)

\bibitem[Jones et al.(2014)]{Jonesetal2014} Jones, D., Boffin, H.~M.~J.,
Miszalski, B., Wesson, R., Corradi, R.~L.~M., \& Tyndall, A. A.\
2014, \aap, 562, A89

\bibitem[Jones et al.(2015)]{Jonesetal2015} Jones, D., Boffin, H.~M.~J.,
Rodr{\'{\i}}guez-Gil, P., Wesson, R., Corradi, R.~L.~M.,
Miszalski, B., \& Mohamed, S.\ 2015, \aap, 580, A19

\bibitem[Kashi \& Soker(2015)]{KashiSoker2015} Kashi, A., \& Soker, N.\ 2015, arXiv:1508.00004

\bibitem[Lagadec et al.(2011)]{Lagadecetal2011} Lagadec, E., Verhoelst,
T., M{\'e}karnia, D., et al.\ 2011, \mnras, 417, 32

\bibitem[Lombardi et al.(2006)]{Lombardi2006} Lombardi, J.~C., Jr., Proulx, Z.~F., Dooley, K.~L., Theriault, E.~M., Ivanova, N., \& Rasio, F.~A.\ 2006, \apj, 640, 441

\bibitem[Mampaso et al.(2006)]{Mampasoetal2006} Mampaso, A., Corradi, R.~L.~M.,
Viironen, K., et al.\ 2006, \aap, 458, 203

\bibitem[Manchado et al.(1996)]{Manchadoetal1996} Manchado, A., Guerrero, M.~A., Stanghellini, L., \& Serra-Ricart, M.\ 1996, The
 IAC morphological catalog of northern Galactic planetary nebulae, Publisher: La Laguna, Spain: Instituto de Astrofisica de Canarias
 (IAC), 1996, Foreword by Stuart R.~Pottasch, ISBN: 8492180609,

\bibitem[Manick et al.(2015)]{Manicketal2015} Manick, R., Miszalski,
B., \& McBride, V.\ 2015, \mnras, 448, 1789

\bibitem[Mardling \& Aarseth(2001)]{MardlingAarseth2001} Mardling, R.~A., \&
Aarseth, S.~J.\ 2001, \mnras, 321, 398

\bibitem[Mart{\'{\i}}nez Gonz{\'a}lez et al.(2015)]{Martinezetal2015} Mart{\'{\i}}nez Gonz{\'a}lez,
M.~J., Asensio Ramos, A., Manso Sainz, R., Corradi, R.~L.~M., \&
Leone, F.\ 2015, \aap, 574, A16

\bibitem[Meaburn et al.(2005)]{Meaburnetal2005} Meaburn, J., L{\'o}pez,
J.~A., Steffen, W., Graham, M.~F., \& Holloway, A.~J.\ 2005, \aj,
130, 2303

\bibitem[Miszalski et al.(2015)]{Miszalskietal2015} Miszalski, B.,
Manick, R., \& McBride, V.\ 2015, in ``Physics of Evolved Stars
2015 - A conference dedicated to the memory of Olivier Chesneau'',
in press (arXiv:1507.07707)

\bibitem[Mo{\v c}nik et al.(2015)]{Mocniketal2015} Mo{\v c}nik, T.,
Lloyd, M., Pollacco, D., \& Street, R.~A.\ 2015, \mnras, 451, 870

\bibitem[Moe \& De Marco(2006)]{MoeDeMarco2006} Moe, M., \& De Marco, O.\
2006, \apj, 650, 916

\bibitem[Monard(2008)]{Monard2008} Monard, L.~A.~G.\ 2008, \iaucirc, 8946, 1

\bibitem[Montez et al.(2015)]{Montezetal2015} Montez, R., Jr.,
Kastner, J.~H., Balick, B., et al.\ 2015, \apj, 800, 8

\bibitem[Muthu \& Anandarao(2003)]{MuthuAnandarao2003} Muthu, C., \& Anandarao,
B.~G.\ 2003, \aj, 126, 2963

\bibitem[Nordhaus \& Blackman(2006)]{NordhausBlackman2006} Nordhaus, J., \&
Blackman, E.~G.\ 2006, \mnras, 370, 2004

\bibitem[Otsuka et al.(2014)]{Otsukaetal2014} Otsuka, M., Kemper, F.,
Cami, J., Peeters, E., \& Bernard-Salas, J.\ 2014, \mnras, 437,
2577

\bibitem[Palen et al.(2002)]{Palenetal2002} Palen, S., Balick, B.,
Hajian, A.~R., Terzian, Y., Bond, H.~E., \& Panagia, N.\ 2002,
\aj, 123, 2666

\bibitem[Passy et al.(2011)]{Passy2011} Passy, J.-C., De Marco, O., Fryer, C.~L., et al.\ 2012, \apj, 744, 52

\bibitem[Passy et al.(2012)]{Passyetal2012} Passy, J.-C., De Marco, O., Fryer, C.~L., et al.\ 2012, \apj, 744, 52

\bibitem[Pottasch et al.(2009)]{Pottaschetal2009} Pottasch, S.~R., Bernard-Salas,
J., \& Roellig, T.~L.\ 2009, \aap, 499, 249

\bibitem[Ransom et al.(2014)]{Ransometal2014} Ransom, S.~M., Stairs, I.~H., Archibald, A.~M., et al.\ 2014, \nat, 505, 520

\bibitem[Ricker \& Taam(2012)]{RickerTaam2012} Ricker, P.~M., \& Taam, R.~E.\ 2012, \apj, 746, 74

\bibitem[Sabach \& Soker(2015)]{SabachSoker2015} Sabach, E., \& Soker, N.\ 2015, \mnras, 450, 1716

\bibitem[Sahai et al.(2011)]{Sahaietal2011} Sahai, R., Morris, M.~R.,
\& Villar, G.~G.\ 2011, \aj, 141, 134

\bibitem[Sahai et al.(2000)]{Sahaietal2000} Sahai, R., Nyman,
L.-{\AA}., \& Wootten, A.\ 2000, \apj, 543, 880

\bibitem[Sahai \& Trauger(1998)]{SahaiTrauger1998} Sahai, R., \& Trauger,
J.~T.\ 1998, \aj, 116, 1357

\bibitem[Sandquist et al.(1998)]{SandquistTaam1998} Sandquist, E.~L., Taam, R.~E., Chen, X., Bodenheimer, P., \& Burkert, A.\ 1998, \apj, 500, 909

\bibitem[Schwarz et al.(1992)]{Schwarzetal1992} Schwarz, H.~E., Corradi, R.~L.~M.,
\& Melnick, J.\ 1992, \aaps, 96, 23

\bibitem[Soker(1994)]{Soker1994} Soker, N.\ 1994, \mnras, 270, 774

\bibitem[Soker(1996)]{Soker1996} Soker, N.\ 1996, \mnras, 283, 1405

\bibitem[Soker(1998)]{Soker1998} Soker, N.\ 1998, \apj, 496, 833

\bibitem[Soker(2002)]{Soker2002} Soker, N.\ 2002, \mnras, 330, 481

\bibitem[Soker(2004)]{Soker2004} Soker, N.\ 2004, \mnras, 350, 1366

\bibitem[Soker(2013)]{Soker2013} Soker, N.\ 2013, \na, 18, 18

\bibitem[Soker(2014)]{Soker2014} Soker, N.\ 2014, arXiv:1404.5234

\bibitem[Soker(2015)]{Soker2015} Soker, N.\ 2015, \apj, 800, 114  % grazing

\bibitem[Soker \& Harpaz(1992)]{SokerHarpaz1992} Soker, N., \& Harpaz, A.\
1992, \pasp, 104, 923

\bibitem[Soker \& Kashi(2012)]{SokerKashi2012} Soker, N., \& Kashi, A.\
2012, \apj, 746, 100

\bibitem[Soker \& Subag(2005)]{SokerSubag2005} Soker, N., \& Subag, E.\
2005, \aj, 130, 2717

\bibitem[Soker et al.(1992)]{Sokeretal1992} Soker, N., Zucker, D.~B., \& Balick, B.\ 1992, \aj, 104, 2151

\bibitem[Szyszka et al.(2011)]{Szyszkaetal2011} Szyszka, C., Zijlstra,
A.~A., \& Walsh, J.~R.\ 2011, \mnras, 416, 715

\bibitem[Tauris \& Dewi(2001)]{TaurisDewi2004} Tauris, T.~M., \& Dewi, J.~D.~M.\ 2001, \aap, 369, 170

\bibitem[Tauris \& van den Heuvel(2014)]{TaurisHeuvel2014} Tauris, T.~M., \& van den Heuvel, E.~P.~J.\ 2014, \apjl, 781, L13

\bibitem[Tokovinin et al.(2006)]{Tokovininetal2006} Tokovinin,
A., Thomas, S., Sterzik, M., \& Udry, S.\ 2006, \aap, 450, 681

\bibitem[Tsebrenko \& Soker(2015)]{TsebrenkoSoker2015} Tsebrenko, D., \& Soker, N.\
2015, \mnras, 447, 2568

\bibitem[Van Winckel et al.(2014)]{VanWinckeletal2014} Van Winckel, H., Jorissen, A.,
Exter, K., Raskin, G., Prins, S., Perez Padilla, J., Merges, F.,
Pessemier, W.\ 2014, \aap, 563, L10

\bibitem[V{\'a}zquez et al.(2002)]{Vazquez2002} V{\'a}zquez, R.,
Miranda, L.~F., Torrelles, J.~M., Olguin, L., Benitez, G.,
Rodriguez, L.~F., Lopez, J.~A.\ 2002, \apj, 576, 860

\bibitem[Webbink(1984)]{Webbink1984} Webbink, R.~F.\ 1984, \apj, 277, 355

\bibitem[Zijlstra(2015)]{Zijlstra2015} Zijlstra, A.\ 2015,
Revista Mexicana de Astronom\'ia y Astrof\'isica, accepted for
publication (arXiv:1506.05508)








\end{thebibliography}
\end{document}